\documentclass{article}

\usepackage{framed,multirow}

\usepackage{amssymb}
\usepackage{latexsym}
\usepackage{graphicx}

\usepackage{url}
\usepackage{xcolor}
\definecolor{newcolor}{rgb}{.8,.349,.1}

\begin{document}

\title{What are the true clusters?}

\author{Christian Hennig,\\
 Department of Statistical Science,\\ 
University College London,\\
Gower Street,\\ 
London WC1E 6BT,\\ 
United Kingdom.\\
  Tel.: +44-020-7679-1698;\\
email: c.hennig@ucl.ac.uk}


\maketitle

\begin{abstract}
Constructivist philosophy and Hasok Chang's active scientific realism
are used 
to argue that the idea of ``truth'' in cluster analysis depends on the
context and the clustering aims. Different characteristics of clusterings are
required in different situations. Researchers should be explicit about on what
requirements and what idea of ``true clusters'' their research is based, 
because clustering becomes scientific not through uniqueness but 
through transparent and open communication. The idea of ``natural kinds''
is a human construct, but it highlights the human experience that the
reality outside the observer's control seems to make certain 
distinctions between categories inevitable. 

Various desirable characteristics of clusterings and various approaches to
define a context-dependent truth are listed, and I discuss what impact these
ideas can have on the comparison of clustering methods, 
and the choice of a clustering methods and related decisions in practice.
~\\~\\
{\bf MCS:} 03A05, 62H30, 91C20
\end{abstract}



\section{Introduction}
Cluster analysis is about finding groups in a set of objects, which are characterized by data that can take various forms such as values of variables, dissimilarities or weighted edges in a graph. The groups may form a partition of the object set, but they may also be overlapping or non-exhaustive. Group memberships may be crisp or fuzzy. Some of the discussion here was written with crisp partitions in mind, some apply to Euclidean space or a given dissimilarity measure, but most thoughts are more general. Cluster analysis is used in many 
different areas with many different aims (see Section \ref{saims} 
for examples). Researchers who apply cluster analysis in practice often want
to know whether the clusters that they find are truly meaningful in the sense
that they correspond to a real underlying grouping. Researchers in the field 
of cluster analysis are interested in whether and which methods are better
at finding the true clusters correctly. In most cluster analysis literature, 
however, explanations of what ``true'' or ``real'' clusters
are, are rather hand-waving. It is widely acknowledged that there is no agreed 
definition of what a cluster is, and in the majority of papers in which new
cluster analysis methods are proposed, the authors do not give a general and 
formal definition of what the ``true clusters'' are that their method is 
supposed to find. 

There is a good reason why there is no generally accepted 
unique definition of true clusters. In different applications, cluster analysis
is used with different aims, and the researchers have different ideas of what
should make the objects belong together that are in the same cluster.
The term ``cluster'' does not mean the same to all researchers in all
situations. This is acknowledged in general overviews and books about
cluster analysis, but seems to be ignored by many authors of specialist work
who try to convince readers that a certain 
method is best for finding the ``true/natural/real'' clusters. Even where
it is acknowledged, this often takes the form of a ``general health warning'',
and consequences regarding the selection and comparison 
of methods and the interpretation of results are rarely spelled out. Is it 
possible to escape the alternative to either make the hardly justifiable 
assumption that there is a unique ``true/natural/real'' clustering against 
which the quality of cluster analysis methods can be objectively assessed, 
or to think that cluster analysis
is somehow arbitrary and ``more of an art than a science'' (\cite{LuWiGu12})? 

My perspective is that of a statistician with expertise in 
cluster analysis and a strong interest in the philosophical 
background of statistics and data analysis. The aim of this paper is to offer
a philosophically informed attitude toward the problem of 
choosing, assessing and interpreting cluster analysis methods and clusterings. 
A key idea is that, given that it depends on the context and
clustering aim what a ``good'' clustering is, researchers need to 
characterize what kind of clusters are required for a given real clustering
problem, and what kind of clusters the different clustering methods are good
at finding, or in other words, in what problem-specific ``truth'' the 
researchers
are interested. Similar ideas have recently been discussed in \cite{AckBDLok10}
and \cite{LuWiGu12}. The present paper can be seen as contributing to 
the research program sketched in those papers, but also as 
enrichening their perspective by adding further philosophical and 
statistical considerations.

In Section \ref{sphilo} I will sketch the philosophical basis of the 
present paper, which complements constructivism with Hasok Chang's 
pluralist active
scientific realism, and I will discuss the concepts of ``natural
kinds'' and ``categorization''. Section \ref{saims} lists 
and discusses various context-dependent
clustering aims. Section \ref{sdef} is about how ``true'' clusters could be 
defined so that they can be used for comparing and assessing different 
clustering methods. Section \ref{sprac} discusses some practical consequences.

\section{Philosophical background}\label{sphilo}
\subsection{Constructivism and science}\label{sconstr}
In the present paper I focus on the question what clusters are ``true'' and/or
``real''. Truth and reality, and to what extent they can be observed, are 
controversial issues in philosophy. My starting point in this respect is my 
constructivist philosophy of mathematical modeling as outlined in 
\cite{Hennig10}, which is connected to 
radical constructivism (\cite{Glasersfeld95})
and social constructionism (\cite{Shotter93}). Radical constructivism is 
based on the idea that the perception and world-view of human beings 
can be interpreted as a construction by the body and the brain of the 
individual, which is seen as a self-organizing system. Social constructionism 
focuses on the construction of a common world-view of social systems by means 
of communication. ``Construction'' refers to the activity of the body, the 
brain, and communicative activity within social systems, setting up
perceptions and world-views. Construction is largely unconscious or 
semi-conscious, and is not arbitrary but subject to constraints. It is not
claimed that individuals or social systems are free to construct any arbitrary
perception or world-view. Experience tells us that perception is
rather severely constrained and shaped by what we perceive to be a reality 
outside of ourselves. 

I distinguish 
observer-independent reality, personal reality and social reality. 
The observer-independent reality is only accessible to humans 
by observation, which means that there is no way to make sure which of its 
features are really observer-independent, but it is usually perceived as the 
source of constraints for personal and social constructs. The perceptions of 
individuals, together with their
thoughts and feelings, make up their personal reality. Part of most personal 
and social realities is the belief that much personal perception 
represents or reflects the observer-independent reality. This belief is 
normally based on the experience of consistency between different sensory 
perceptions, 
at different times and from different positions, and on the 
confirmation of the existence of many of the perceived items by 
communication with others. It is therefore the result of active 
accommodation of perceptions. 

Social reality is made up by communication between individuals. It
is carried by social systems, which may overlap
and may partly lack clear borderlines, although some social systems such as 
formal mathematics are rather clearly delimited. Personal and social realities
influence each other. According to the point of view taken here,
science is a social attempt to construct a
consensual and stable view of the world, which can be shared by everyone and 
is open to criticism and scrutiny in free exchange. In this sense, science 
aims at a view that is as independent as possible of the individual observer,
and is therefore connected to a traditional realist view, according to which
science aims at finding out the truth about observer-independent reality.
But constructivists are pessimistic regarding an 
observer-independent access to reality, and assess the success of science 
based on stability, agreement and pragmatic use instead of referring to
objective truth. A scientific world-view with which 
constructivists can agree needs to acknowledge the existence and
legitimacy of diverse personal and social realities and is therefore 
inherently pluralist. A tension between a drive for unification
and general agreement and a necessity to allow space for diverse realities in
order to allow for criticism and creative progress is an essential 
implication of the scientific idea.
Central tools of science are mathematics, which aims at
setting up and exploring concepts that are clear and well defined 
independently of the different personal and social points of view and at 
statements about which absolute agreement is possible, and 
measurement, which unifies observations of reality in a way that they can be
processed by mathematical means. 

Constructivism is often accused of
denying the existence of the observer-independent reality altogether by
calling it ``a construct'', but actually, being
as stable and ubiquitous a construct as the observer-independent reality
seems to be in most personal and social realities, 
it is as real as anything can get in constructivism. 

\subsection{Active scientific realism}
Although constructivism is often interpreted as anti-realist, I complement my
constructivist view here by the ``active scientific realism'' introduced
by Hasok Chang (\cite{Chang12}). In the abstract of his Chapter 4, Chang writes:
{\it ``I take reality as whatever is not subject to one’s will, and knowledge as
an ability to act without being frustrated by resistance from reality. 
This perspective allows an optimistic rendition of the pessimistic 
induction, which celebrates the fact that we can be successful in science 
without even knowing the truth. The standard realist argument from 
success to truth is shown to be ill-defined and flawed. I also reconsider 
what it means for science to be ``mature'', and identify humility rather
than hubris as the proper basis of maturity. The active realist ideal is 
not truth or certainty, but a continual and pluralistic pursuit of 
knowledge.''} Chang's use of the term ``reality'' refers to what is vital 
for the success of the scientific idea, namely to confront scientific
work continually with the observed realities that individuals and social 
systems experience as outside their control. In agreement with my 
constructivist view, active scientific realism values
a plurality of perspectives. The term ``truth'' is used in both
\cite{Chang12} and the constructive literature as a relative concept 
``internal to systems of practice''. For example, within the mathematical 
formal system, ``truth'' is a rather unproblematic concept due to the clear
rules by which it can be ensured, whereas the truth-value of the statement 
``the German Democratic Republic was a democracy'' depends
on which characteristics of a political system are taken as essential for
being a democracy, which differs between social systems. 

The emphasis of the strong role of communication and language is an 
aspect that constructivism adds to active scientific realism. In this respect
I follow \cite{Fleck79}, a pioneer work regarding the role of 
communication and social systems (``thought collectives'') 
for scientific knowledge. Fleck showed how scientific facts are shaped by 
the specific way how collectives of scientists conceptualize their field. 

\subsection{Natural kinds}
``Natural kinds'' in philosophy
refer to the idea that there
are some ``naturally'' separated classes in observer-independent reality,
which, for traditional realists, correspond to ``true clusters''. For example,
biological species and chemical elements are considered
as candidates for being natural kinds (\cite{sep-natural-kinds}). There is 
much controversy about what constitutes natural kinds (e.g., common
properties, behaving homogeneously according to natural laws). The 
concept runs counter to the constructivist view that what is perceived as 
``kinds'' is constructed by human activity and language and depends on 
the conditions of observation and practice of living of the observers. For 
such reasons, for example
\cite{Goodman78} rejected the term ``natural'' for kinds. 
\cite{Hacking91} argued that ``natural kinds'' should refer to 
kinds that are connected to human activity and utility, which allows for
non-uniform and more pluralist kinds. According to him, the concept links
a nominalist inclination with a traditional realist view of ``nature''.
He also suggested that many
classes that can be seen as natural in some sense are not
``natural kinds'', and that this term may be reserved for a few
very special kinds. 

I agree with Goodman that the term ``natural'' is not helpful, at least if it
is used in order to suggest that some categorizations have a special
authority by matching observer-independent reality. What is valuable about the
concept of ``natural kinds'' is that it describes a human experience that
certain categorizations seem impossible to escape when confronted with Chang's 
``reality outside our control''. Such an experience always has to be framed by 
the make-up of the personal and social realities that are involved, it may 
change, and controversy persists even about central candidates for 
natural kinds such as biological species (\cite{Hausdorf11}) and chemical
elements (\cite{Chang12}). Still, it highlights that when following 
an active scientific realist agenda, phenomena should not be lumped
arbitrarily into classes, but that scientific observation should be used to 
guide classification in a stable way that should aim at general agreement;
by which I mean agreement about the legitimacy and use of the classification
as opposed to its uniqueness. 

\subsection{Categorization}
From the constructivist point of view, although we experience 
``reality outside our control'', the categorization of its phenomena is a 
constructive human activity, and any idea of ``true'' or ``really meaningful'' 
categories is located in personal and social reality. In order to understand
such an idea it therefore seems promising to look at work in
cognitive science about human categorization. \cite{MeHaMiTh93} review 
cognitive theories of categorization with a view to connecting them to 
inductive data analysis including clustering. Although no explicitly pluralist
position is taken in that book, the various presented theories seem to apply to 
different kinds of categories used by human beings in different circumstances.
Many of these theories correspond to formal approaches to cluster analysis, for
example that categorization can be based on defining features, prototypes and 
exemplars, or family resemblance (similarity). From a constructivist 
perspective, \cite{Foerster81} saw ``objects'' in human perception as
eigenvalues (fixed points) of recursive coordinations of actions, which has a
reflection in self-organizing clustering algorithms. 
Because of the exchange between cognitive science and artificial intelligence 
research, this should not be surprising. 
However, formal and algorithmic views 
of categories have strong limitations, and it has been pointed out that 
in order to understand human categorization, context such as the conditions
of the human body, a metaphorical or theoretical framework in which a category
is embedded (\cite{Lakoff87}, Chapter 7 of \cite{MeHaMiTh93}) and the 
ever-changing social and communicative environment (\cite{Shotter93}) need 
to be taken into account. 

Another line of research concerns intuitive clustering by humans of two
dimensional point clouds, regardless of the meaning of the points, see
\cite{SaSa05,Lew09}, with mixed results in the sense that there are
predominant stategies such as looking for high density areas and 
for shapes of similar kinds (``model fitting''), but there is also
considerable variation, although \cite{LADS12} argue that humans and 
particularly experts are more consistent in assessing clusterings 
than existing cluster validation indexes.

Overall, categorization seems to work in rather pluralist and 
context-dependent ways, as is also acknowledged in more recent publications
on categorization
(\cite{AshMad05,RSM12}). It may be controversial to what extent 
cluster analysis methods are meant to reflect human categorization. One could
argue that ``true clusters'' should have a more scientific and well-defined
character than the concepts that humans normally use. Furthermore, 
clustering often aims at finding categories that are thought of as determined 
by unobserved features, which differs from forming
categories from what is observed.
The theories discussed in this section are relevant in artificial intelligence 
applications where the aim is to simulate human categorization, and they 
can also inspire methodological ideas in
clustering, but their potential to define ``true clusters'' as targets 
for data analysis is limited.

\section{Clustering aims and cluster concepts}\label{saims}
\subsection{A list of aims of clustering}
That there is no generally
accepted definition of a cluster is not surprising, given the many
different aims for which clusterings are used. Here are some examples:
\begin{itemize}
\item delimitation of species of plants or animals in biology, 
\item medical classification of diseases,
\item discovery and segmentation of settlements and periods in archeology,
\item image segmentation and object recognition,
\item social stratification,
\item market segmentation,
\item efficient organization of data bases for search queries.
\end{itemize}
There are also quite general tasks for which clustering is applied in many
subject areas:
\begin{itemize}
\item exploratory data analysis looking for ``interesting patterns'' without 
prescribing any specific interpretation, potentially creating new research
questions and hypotheses,
\item information reduction and structuring of sets of entities from 
any subject 
area for simplification, effective communication, or effective 
access/action such as complexity reduction for further data analysis,
or classification systems,
\item investigating the correspondence of a clustering in specific data
with other groupings or characteristics, either 
hypothesized
or derived from other data.  
\end{itemize} 
Depending on the application, it may differ a lot what is meant by a 
``cluster'', and cluster definition and
methodology have to be adapted to the specific aim of clustering in
the application of interest.

\subsection{Realist and constructive aims of clustering}
A key distinction can be made between ``realist'' and ``constructive''
aims of clustering. Realist aims
concern the discovery of some meaningful real structure (referring to what
is experienced as ``reality outside our control'', see Section
\ref{sphilo}). Constructive aims refer to the researchers' 
intention to split up
the data into clusters for pragmatic reasons, regardless of whether there is 
some essential real difference between the resulting groups. The
connection between 
``realist'' and ``constructive'' clustering aims and
realist and constructivist philosophy is not straightforward. 
Nothing stops a realist from being interested in data compression 
and from therefore having a constructive clustering aim. On the other hand,
a constructivist can legitimately be interested in realist clustering aims,
although she would maintain that the idea of clusters that are 
real and meaningful in the observer-independent reality
is a personal and social construct. 

The distinction between realist and constructive clustering aims is 
not clear cut. As follows from Section
\ref{sphilo}, researchers with realist clustering aims  
should not hope that the data alone reveals real structure; constructive 
impact of the researchers is needed to decide what counts as real.

The key issue in realist clustering is how the real 
structure the researchers are interested in is connected to the available 
data. This requires subject matter knowledge and 
decisions by the researchers.   
``Real structure'' is often understood as the existence of an unobserved 
categorical variable, the values of which define the ``true'' clusters. 
But neither can it be taken fur 
granted that the categories of such a variable
are the only existing ones that could qualify
as ``real clusters'', nor do such categories necessarily correspond to data
analytic clusters. For example, male/female is a meaningful 
categorization of human beings, but there may not be a significant 
difference between men and women regarding the results of a certain attitude 
survey, let alone separated clusters corresponding to sex. Usually the objects
represented in a dataset can be partitioned into real categories in many
ways. Also, different cluster analysis methods will produce different 
clusterings, which may correspond to patterns seen as  ``real'' 
in potentially different ways. This means that in order to decide about
appropriate cluster analysis methodology, researchers need to think about
what data analytic characteristics the clusters they are aiming at are supposed
to have. I call this the ``cluster concept'' of interest in a study.

The real patterns of interest may be more or less closely connected to the 
available data. For example, in biological species delimitation, the concept of
a species is often defined in terms of interbreeding (there is some
controversy, see \cite{Hausdorf11}). But interbreeding 
patterns are not usually available as data. Species are nowadays usually 
delimited by use of genetic data, but in the past, and occasionally in 
the present in exploratory analyses, species were seen as the source of a 
grouping in phenotype data. In any case, the researchers need an 
idea about how true distinctions between species are connected to 
patterns in the data. Regarding genetic data, knowledge needs
to be used about what kind of similarity arises from persistent genetic 
exchange inside a species, and what kind of separation arises between distinct
species. There may be subgroups of individuals in a species 
between which there is little 
actual interbreeding (potential interbreeding suffices for forming a
species), e.g., geographically separated groups, and 
consequently not as much genetic similarity as one would naively expect.
Furthermore there are various levels of classification in biology, such as
families and genii above and subspecies below the level of species, so that 
data analytic clusters may be found at several levels, and the researchers
may need to specify more precisely how much similarity within and separation
between clusters is required for species.

Such knowledge needs to be reflected in chioce of
the cluster analysis method. 
E.g., species may be very heterogeneous regarding geographical
distribution and size, and therefore a clustering method that 
penalizes large within-cluster distances too heavily
such as $k$-means or complete linkage is inappropriate. 

In some cases, the data are more directly connected to the cluster definition.
In species delimitation, there may be interbreeding data, in which 
case researchers can specify the requirements of a clustering more directly.
This may imply graph theoretic clustering methods and a specification of 
how much connectedness is required within clusters, although such decisions
can often not be made precise because of missing information arising from 
sampling of individuals, missing data etc. On the other hand, the connection
between the cluster definition and the data may be less close, as in the case
of phenotype data used for delimiting species, in which case some 
speculation is needed in order to decide
what kind of clustering method may produce something useful.
 
In many situations different groupings can be interpreted as real, depending on
the focus of the researchers. E.g., social classes can be defined in 
various ways. Marx made ownership of means of production the major 
defining characteristic of different classes, 
but social classes can also be defined 
by looking at patterns of contact, or occupation, or 
education, or wealth, or by a mixture of these (\cite{HenLi13}). 
In this case, a major issue 
for data clustering is the selection of the appropriate variables and 
measurements, which implicitly defines what kinds of social classes can be 
found. 

The example of social stratification illustrates that there is a gradual
transition rather than a clear cut between realist and constructive 
clustering aims. According to some views (such as the Marxist one) 
social classes are an essential and real characteristic of society, but 
according
to other views, in many societies there is no clear delimitation between
supposedly ``real'' social classes, despite 
the existence of real inequality. 
Social classes can then still be used as a convenient tool for
structuring the inequality. 

Regarding constructive clustering aims, it is obvious that researchers need
to decide about the desired ``cluster concept'', i.e., about
the characteristics that their clusters should have. This 
needs to be connected to the 
practical use that is intended to be made of the clusters. 

Where the primary clustering aim is constructive, realist clustering may still
be of interest. If indeed some real grouping structure is
manifest in the data, many constructive aims will be served well by
having this structure reflected in the clustering. E.g., market 
segmentation may be useful regardless of whether there are really 
meaningfully separated groups in the data, but it is relevant
to find them if they exist.

\subsection{Desirable characteristics of clusters} \label{sdesir}
Here is a list of potential characteristics of clusters that may be desired,
and that can be checked using the available data. Several of these are 
related with the ``formal categorization principles'' listed in Section 
14.2.2.1 of \cite{MeHaMiTh93}.
\begin{enumerate}
\item Within-cluster dissimilarities should be small. 
\item Between-cluster dissimilarities should be large.
\item Clusters should be fitted well by certain homogeneous 
probability models such as the Gaussian or a uniform distribution on 
a convex set, or by linear, time series or spatial process models.  
\item Members of a cluster should be well represented by its centroid.
\item The dissimilarity matrix of the data should be well represented by the
clustering (i.e., by the ultrametric induced by a dendrogram, or by defining
a binary metric ``in same cluster/in different clusters''). 
\item Clusters should be stable.
\item Clusters should correspond to connected areas in data space with high 
density. 
\item The areas in data space corresponding to clusters should have certain
characteristics (such as being convex or linear).
\item It should be possible to characterize the clusters using a small number
of variables.
\item Clusters should correspond well to an externally given partition or 
values of one or more variables that were not used for computing the 
clustering. 
\item Features should be approximately independent within clusters.
\item All clusters should have roughly the same size.
\item The number of clusters should be low.
\end{enumerate}
When trying to measure these characteristics, they have to be made more precise,
and in some cases it matters a lot how exactly they are defined. Take no. 1,
for example. This may mean that all within-cluster dissimilarities 
should be small (i.e., their maximum, as required by complete 
linkage clustering), or their average, or a high quantile of 
them. These requirements may look similar at first sight but are very 
different, e.g., regarding the integration of outliers in clusters. 
Having large
between-cluster dissimilarities may emphasize gaps by looking at the smallest
dissimilarities between two clusters, or it may rather mean that the 
cluster centroids are well distributed in data space. 
 As another example, stability can refer to sampling other data from the same
population (this may play a priviliged role in hypothesis driven repeated experiments aiming at reproducible results, which is often identified with the scientific method), to adding ``noise'', or to comparing results from different 
clustering algorithms (\cite{BDVLP06}). 

Some of these characteristics conflict with others in some datasets. E.g., 
connected areas with high density may include very large distances,
and may have shapes that are undesired in specific applications
(e.g., non-convex). Representing 
objects by centroids well may require some clusters with little or no 
gap between them. Stability is often easier to achieve with fewer clusters
than required in situations where clusters need to be very homogeneous

Deciding about such characteristics is the key to linking the clustering 
aim to an appropriate 
clustering method. E.g., if a database of images should be clustered 
so that users can be shown a single image to represent a cluster, 
centroid representation is 
most important. Useful market segments need to be addressed by 
non-statisticians and should therefore normally be represented by few
variables, on which dissimilarities between members should be low. 
Section \ref{sprac} outlines how the listed characteristics can help with the
selection of a clustering method in practice. 

The idea of listing potentially desirable characteristics
of clusterings for helping with the selection of clustering methods
is central also to \cite{AckBDLok10}, but the axiomatic
characteristics listed there are strikingly different from the present list.
As necessary for the theoretical analysis, the characteristics in 
\cite{AckBDLok10} are formal. One reason for the differences may be that
the aim of the authors was to prove general theorems, and therefore they
went for characteristics that make such theorems possible. \cite{ABDBL12} 
and \cite{ABDSL13} investigated cluster analysis approaches with
respect to further formal characteristics, which are related to some of
the characteristics listed above. Ultimately, characteristics 
need to be formalized to be used in practical analyses, in which case at
least some of them (distance to centroids, quality of representation of the
data and fit by probability models) also serve to measure information loss 
through clustering. 
Similar considerations can be found in 
\cite{LuWiGu12}, which are closer to the present approach, but somewhat 
less detailed. Ultimately, the characteristics listed here need to be 
formalized, too, to be used in practical analyses. 

\section{Definitions of true clusters}\label{sdef}
There is no agreed definition of what true clusters are in reality, but 
mathematical formalism allows to give a clear definition (a mathematical 
model) of true clusters based on mathematical objects. In different
situations, different kinds of clusters 
are of interest, and a mathematical definition of true clusters
cannot be unique. It is necessarily idealized and abstract, and discrepancies 
between such a definition and the more complex and informal 
ideas that researchers have about reality should not be suppressed
(see \cite{Hennig10} for a constructivist view of mathematical models).

Still, an explicit formal definition of true clusters has important benefits.
It communicates the cluster concept in a specific setup in a clear way, and
it provides a transparent framework for comparing methods. It may also
stimulate the development of new methodology. In the literature on clustering
methods, clear definitions of the specific clustering problem to be solved 
are often missing, probably because authors feel that such definitions could not
properly cover the clustering problem in general. But this means that a chance
is missed to clarify the understanding of what kind of problem a method is 
good or not so good for. 

For every formal definition there need to be arguments why it formalizes 
a reasonable cluster concept researchers could
be interested in, so it needs to be related to desirable characteristics of
clusters. 
Definitions of true clusters can be based on the data, which are 
measurements that therefore ``live'' in the system of mathematical formalism.
This is only appropriate if what makes a certain subset of the
data a true cluster according to the researchers can indeed be defined
from the data alone. For realist clustering aims, true clusters need to be 
defined based on a certain truth ``behind'' the data. There are two 
possibilities for doing this. Firstly, one could assume that in the 
``mathematical world'' there is true clustering information for all
observations, which is available in principle but not used by the clustering 
method. Secondly, one could assume that the data are generated by a
true probability model, and then define the truth in terms of this model. 

\subsection{Definitions based on the data alone} \label{sdefdata}
Let ${\bf x}_1,\ldots,{\bf x}_n$ be $n$ observations in $\mathbb{R}^p$. 
$k$-means clustering is defined by choosing $k$ cluster mean
vectors ${\bf a}_1,\ldots,{\bf a}_k$ and a cluster assignment 
function $\gamma: \{1,\ldots,n\}\mapsto \{1,\ldots,k\}$ so that
$\sum_{i=1}^n  \|{\bf x}_i-{\bf a}_{\gamma(i)}\|^2$ is minimized. The solution
of this problem could be called ``the true clustering''. 

Is this appropriate? It could be, namely if the real aim is to find a 
clustering with $k$ clusters in which all observations are represented optimally
(in the sense of averaging the squared Euclidean distance) by the centroid
of the cluster to which they are assigned. On the other hand, if in the 
situation of interest clusters should rather correspond to high-density
regions, clusters defined as ``true'' by $k$-means can be inappropriate, see
Figure 2 for an example. Note also that for defining true clusters
according to the $k$-means criterion, $k$ has to be assumed to be known.

Is such a definition helpful? 
If the $k$-means objective function is used to define the 
true clusters, obviously $k$-means clustering is the best clustering method,
and this may look tautological, although it is still of interest to 
investigate what extent
different algorithms are successful for minimizing the objective function.

In principle, if the objective function that defines a clustering method 
corresponds exactly to the loss function of the practical problem for which a clustering is 
required, there is no point to look for other clustering methods. The same
holds for methods that are not defined by optimizing an objective function 
but, e.g., are stable states reached by an algorithm, as long as this is 
for solving a practical problem properly formalized by the algorithm. 
In this sense,
most clustering methods implicitly define their own truth. A practical 
implication is that the definition of a clustering method often
gives strong information about what kind of clustering problem the method is 
good for. 

However, in most clustering applications the aims of clustering do not directly
translate into a specific cluster analysis method, be it through matching
the practical ``loss'' with the method's  objective function or otherwise. In
general, the choice of the the practical ``loss'' and therefore the
objective function or more generally 
the clustering principle needs to be supported by 
validation techniques and background information.

In some other situations it is possible to define a clustering problem based
on the data alone without corresponding directly to any available 
clustering method. An example for this is the optimal
approximation of the distance matrix of the data by an ultrametric induced 
by a dendrogram produced by a hierarchical clustering method. Another approach 
would be the definition of an aim-dependent cluster quality index as a 
weighted mean of appropriately scaled statistics measuring cluster 
characteristics as listed in Section \ref{sdesir} (in \cite{LuWiGu12} there
is a related discussion of measuring and optimizing ``usefulness'' of 
clusters). In an implicit manner, internal cluster validation indexes
(\cite{XiongLi14}) such as the average silhouette width attempt to 
aggregate desirable features of clusterings, and ``true clusters'' could be
defined by optimizing them, although such criteria are usually designed 
with the aim of defining a too general notion of cluster quality,
which does not take into account the differences between clustering aims in 
practice. 

If ``truth/quality'' is defined in such a way, one could try to optimize
the cluster quality index directly. This
is often not computationally feasible, and also in some cases 
desirable characteristics need to be combined in other ways than just 
averaging them (for example, one may be interested in constrained 
optima of objective functions, putting an upper bound on within-cluster 
distances). So there is still a place for clustering methods that do not 
directly optimize a quality index. Also, clustering applications in which
the idea of truth refers to the observed data alone are probably a small
minority; particularly it implies that the data cover all objects of interest
and are not only a sample from which the researchers want to generalize.

Some other work explores notions of ``clusterability'' 
of data (\cite{BBV08,AckBD09}), revealing that there are several reasonable
notions that contradict each other in many situations. 

\begin{figure}[t]
\label{f1}
\centering
\includegraphics[width=.45\textwidth]{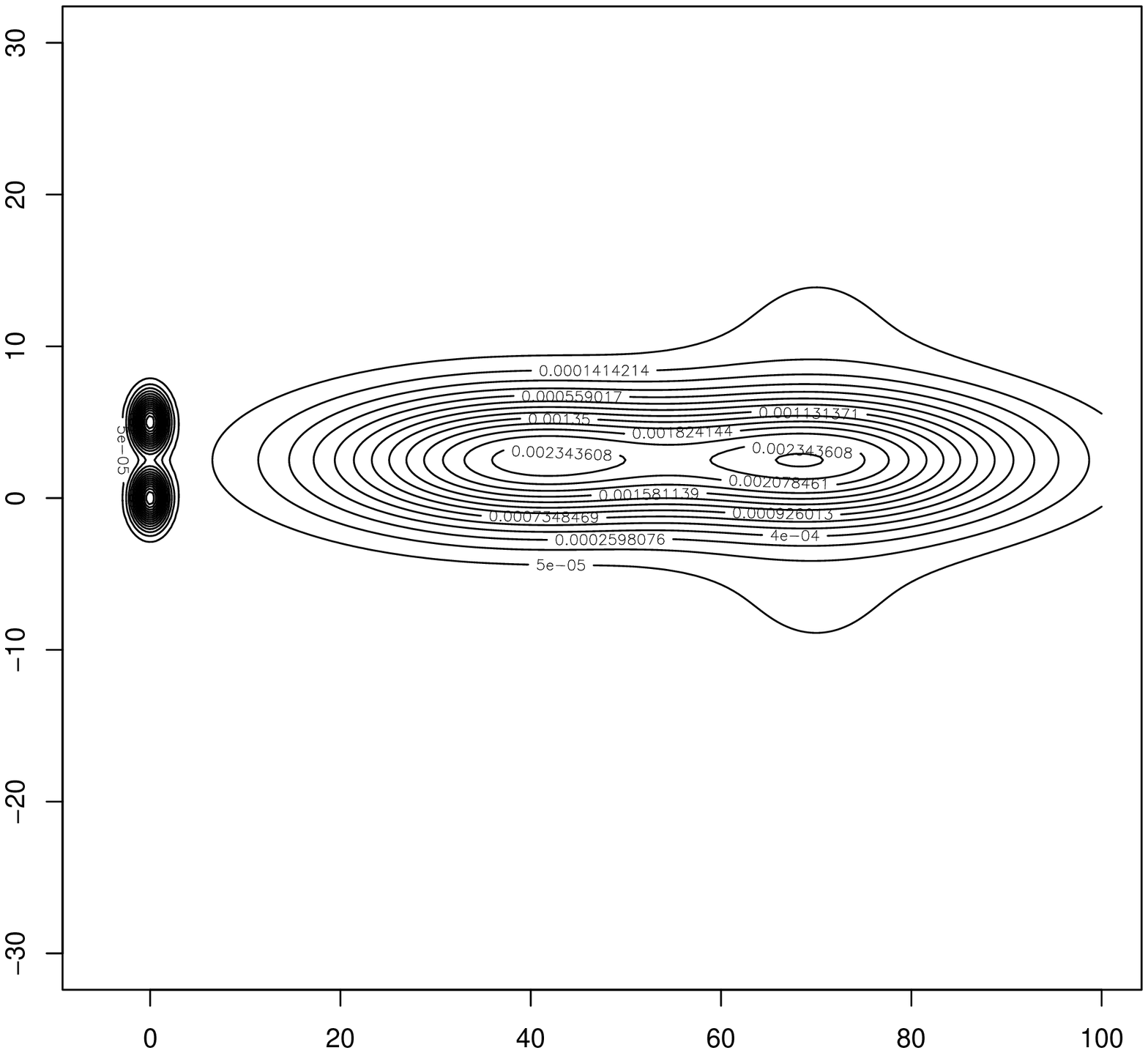}
\includegraphics[width=.45\textwidth]{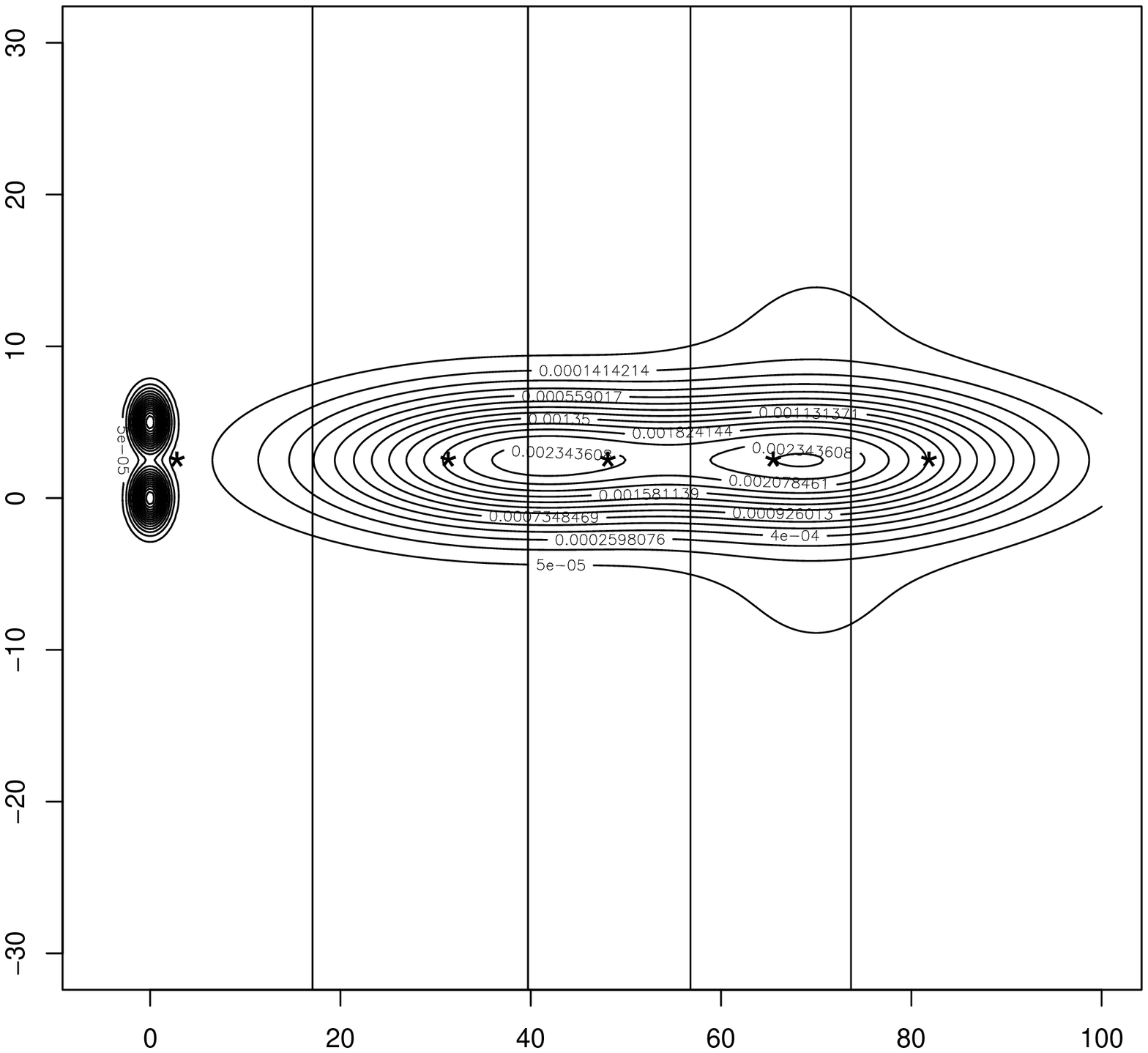}
\caption{Density contour of a mixture of five Gaussian distributions (mean
vectors are $(0,0),\ (0,5),\ (40,2.5),\ (70,2.5)$; there are two components 
centered at $(70,2.5)$ with different covariance matrices). Below:
optimal 5-means partition and mean vectors (asterisks).} 
\end{figure}

\begin{figure}[t]
\label{f2}
\centering
\includegraphics[width=.45\textwidth]{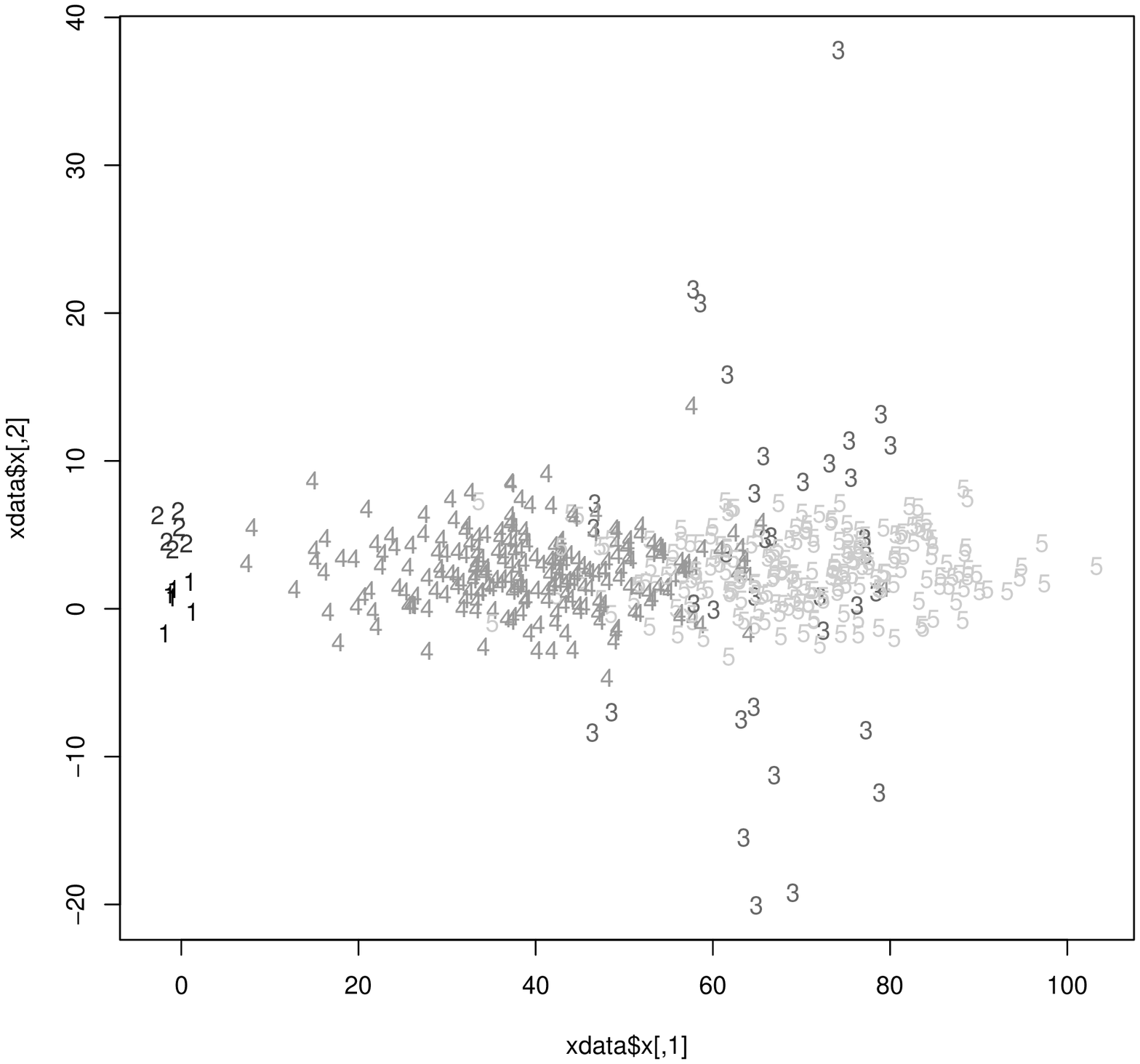}
\includegraphics[width=.45\textwidth]{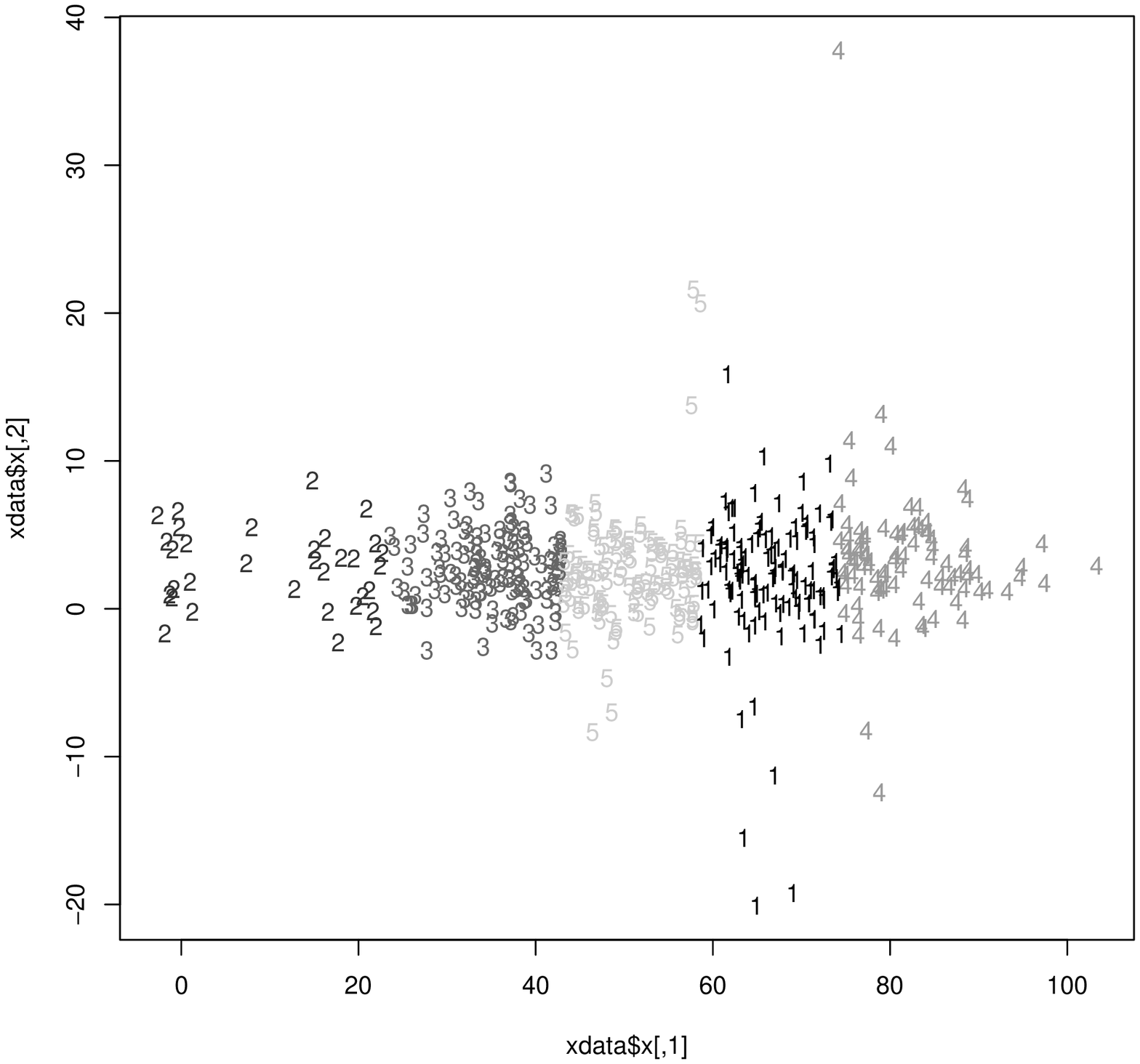}
\caption{Data generated from model in Figure 1, above: mixture 
components from which observations were generated, below: 5-means
clustering.}
\end{figure}

\subsection{Definitions based on external information}\label{sext}
In comparisons of cluster analysis methods in the literature, authors often
use datasets for which there is a given ``true classification''. Often these
are standard examples for supervised classification such as Fisher's famous
Iris dataset in which there are measurements on 150 Iris plants from three 
different subspecies.  Clustering methods can generate clusterings ignoring
the true classification to which they then can be compared. 

This is an artificial situation. In reality cluster analysis is
applied to find clusters that are not yet known. The appeal of this approach
is that realistic datasets can be used and that it is usually 
easy to argue that the true given classes are meaningful. But often
measuring the performance of clustering methods on datasets with given true
classes is not very informative. How informative it is depends on to what 
extent the true classes in such cases are good models for the true clusters
the researcher wants to find in a new dataset with unknown truth. This is hardly
ever discussed. Usually, it is not investigated to what extent the true
given classes have the desired characteristics of clusters in the situation
of interest. There is no guarantee
that true classes from supervised classification problems qualify as ``data 
analytic clusters'' (in the sense of the previous subsection), and it
may not be reasonable to expect a good clustering method to find them. 
Furthermore, there is no guarantee that the given true classes are the only
categorical variable that qualifies for defining true classes; there could be
further (unobserved) variables defining alternative true classes. 

Although such real datasets with given true classes can contribute
to the comparison of clustering methods, the approach seems to be overused 
in the literature, and where it is used, more care is required for exploring
what can be learned for other datasets without known classes from the 
``success'' of certain methods to recover known true classes.

The same applies to the presentation of datasets for which 
authors refer to some ``truth'' without a formal definition, just appealing
to the reader's (usually Euclidean) intuition. E.g., data distributed
on a ball about the origin together with data distributed around a much wider
circle about the origin with a hole in the middle that separates it from
the central ball are often presented as an illustration that ``$k$-means
does not work'', not reproducing the clustering
the authors declare to be true by fiat. This clustering is based on separation, 
but the biggest distances in the dataset occur within a cluster, namely the
wider circle, so this qualifies as ``true cluster'' in some respects but not 
others. Euclidean intuition is irrelevant in a large number of clustering
problems (e.g., with categorical variables or non-Euclidean dissimilarities)
and should not be overrated as reliable indicator of ``truth'' in Euclidean
setups either. Again, such data can be used in a constructive way for 
evaluating clustering methods, but reference to the specific characteristics
of the given true clustering needs to be made.

External information can also be used in other ways to define cluster 
quality (and therefore implicitly the ``true clusters'' by optimizing 
quality). In applications where clustering is used instrumentally for some
other aims of data analysis, for example for data compression in order to 
predict an external variable, different clusterings can be compared according
to quality measures related to the final aim, e.g., prediction quality.

\subsection{Definitions based on probability models}\label{sdpm}
Assuming that data are generated from
probability models is the standard technique for defining true underlying but
unobserved clusters. It can then be investigated by (asymptotic) theory
or systematic simulation whether cluster analysis methods find such 
clusters. There are various approaches to define true clusters based on 
probability models. Most straightforward are mixture models of the form
$f({\bf x})=\sum_{j=1}^k \pi_jf_{\theta_j}({\bf x})$, where data ${\bf x}$ are
assumed to be i.i.d. generated from a distribution with density $f$ with is a 
mixture of parametric densities $f_{\theta_j}$. This 
models that ${\bf x}$ is generated
from mixture component $f_{\theta_j}$ with probability $\pi_j$, and data
can be simulated by simulating the true component memberships first. The usual 
interpretation is that the true clusters correspond to the mixture components.
Clusterings computed from the data ${\bf x}_1,\ldots,{\bf x}_n$ can be 
compared to the true component memberships for simulated data. 

Although such a definition gives researchers a much clearer idea of the 
involved cluster concept than using a given true class for real data,
there are several issues with this approach. 

Firstly, the family of
mixtures of distributions of the form $f_{\theta}$ needs to be identifiable, 
i.e., no two sets of parameters $\{(\pi_1,\theta_1),\ldots,(\pi_k,\theta_k)\}$
should generate the same probability measure. This is fulfilled for most
popular mixture models including Gaussian mixtures. If mixtures are 
considered in full generality of the concept, however, identifiability cannot
be taken for granted. Uniform distributions on connected sets can be pieced 
together from uniform distributions on subsets in different ways. Gaussian
mixtures can be written down as mixtures of truncated Gaussians, which 
are no longer identifiable. 
This indicates that parametric families that generate 
identifiable mixtures are chosen rather for technical reasons than because 
they would be 
particularly qualified for representing a clustering ``truth'' in reality.

Secondly, identifying
clusters with mixture components may intuitively not be justified. 
The parametric family needs to be chosen in such a way that 
the $f_{\theta}$ can indeed
be interpreted as ``cluster shaped'', as prototypical models for clusters
of interest. But two 
parameters $\theta_1$ and $\theta_2$ may be so close to each other that the 
mixture of distributions $\pi_1f_{\theta_1}+\pi_2f_{\theta_2}$ may be unimodal,
and may look so homogeneous that it would be inappropriate to split it up
into two clusters in a real application. Figure 1 shows a density 
contour of a Gaussian mixture with five components but only four modes, two of
which are not separated by a deep density valley. Figure 2 shows some 
data generated from this mixture. It strongly depends on the application 
whether it is appropriate to interpret this distribution as generating five 
clusters. Note that there are very large distances within some of the 
mixture components, and it is hard to argue that the points from component 3 
``belong together''. 
One may wonder whether 
mixtures of homogeneous distributions such as the Gaussian should be 
interpreted as single clusters if their mixture is homogeneous enough, 
which allows for more flexible cluster shapes, but  
violates identifiability and requires the researcher to define under
what conditions mixture components should be merged
(\cite{Hennig10merge}). 

Thirdly, statisticians do not believe that parametric 
probability models hold precisely in reality, but true clusters as
mixture components are only well defined if the mixture model holds precisely.
This problem is worse for mixture models than elsewhere in
parametric statistics, because if data come from a distribution 
with a density $g$ that is
slightly different than $f=\sum_{j=1}^k \pi_jf_{\theta_j}$ with a 
certain $k$, $g$ can (under weak assumptions) be 
approximated arbitrarily well by a mixture $f^+$ of distributions of the
form $f_{\theta}$ with $k^+>k$ mixture components, which means that $g$ 
can be approximated by a distribution with more and potentially quite 
different true clusters, despite being so close to $f$ that 
it would require a very large dataset to tell $f$ and $g$ apart.

Despite such problems, defining true clusters as mixture components 
at least communicates a clear idea of a ``cluster prototype model'', and 
allows tests whether clustering methods recover the true clusters in such
mixtures. Such tests can be expected to favor clustering methods that are
based on parameter estimators (e.g., maximum likelihood (ML)). A more
comprehensive evaluation needs to consider models that are approximately but
not precisely equal to such mixtures, and cases in which the interpretation of
single mixture components as clusters breaks down, e.g., because 
mixtures of several components are homogeneous in some sense.

Alternatively, true clusters could be defined as high density level sets
or attraction areas of density modes of distributions. This requires only
the weaker nonparametric assumption that a density exists. Although this is more
general than the mixture approach and allows for more flexible 
cluster shapes (which may or may not be desired), it does not solve 
all the problems connected to the mixture approach. For every 
distribution $P$ with a density and $k$ modes there are distributions 
without an existing density and distributions with an arbitrarily 
higher number of density modes that are so similar to $P$ 
that they cannot be distinguished by an arbitrarily large amount of data
(\cite{Donoho88}). As
the mixture model approach, the density-based approach does not 
generalize to a full neighborhood of $P$. 

A third approach is to define true clusters through statistical functionals of
distributions. This allows for example to generalize the definition of 
$k$-means to distributions $P$, defining true underlying (unobserved)
$k$-means-type clusters, by defining ${\bf a}_1,\ldots,{\bf a}_k$ and 
$\gamma: \mathbb{R}^p \mapsto \{1,\ldots,k\}$ as minimizers of
$\int  \|{\bf x}-{\bf a}_{\gamma({\bf x})}\|^2 dP({\bf x})$. For some other 
clustering methods (including ML estimation for mixtures)
corresponding notions of truth can be defined in similar 
ways; see Section \ref{sdefdata} for comments on adapting the cluster 
definition to a certain method. The formalization using probability models 
allows the investigation of the asymptotic properties of the methods. E.g., 
\cite{Pollard81} proved the consistency of $k$-means applied to data
as estimator for the $k$-means functional. Such functionals can in principle
be defined for any distribution; a density is not required, but in case
of the $k$-means functional existence of second moments is necessary. 
The $k$-means functional can still vanish or change rapidly in the 
neighborhood of any distribution $P$. \cite{Davies93} argued (for 
linear regression) that statisticians should be interested in estimating
globally defined and continuous functionals of distributions, because only 
such functionals cannot change arbitrarily in the neighborhood of a 
distribution. The clustering problem, though, is inherently discontinuous
in borderline situations where a cluster splits, where the number of clusters 
changes or is misspecified (as far as I know, all currently existing
functional-type definitions of true clusters require the number of clusters
to be fixed). 
 
These different approaches to define the truth illustrate that the clustering
problem does not boil down to estimating the underlying distribution. 
Genuinely different true clusterings can be defined for the same
distribution. The distribution showed in Figure 1 is a mixture of five
Gaussian components, has four density modes and (with appropriate level set 
cutoff is) three high density level-sets. 
The right side shows the true
5-means-type functional partition of the distribution. This may look 
counter-intuitive, and it is important to argue that any definition
of true clusters based on a distribution formalizes a clustering that has
certain desirable characteristics. But in the specific case that researchers
want to find cluster centroids so that observations can be 
represented optimally by the centroids in the $k$-means sense, even such a
counter-intuitive partition can be seen as ``true''.

\subsection{Limitations of formal definitions}

All the definitions listed above have shortcomings. Definitions
based on the data alone do not reflect the idea of an unobservable
underlying truth and of generalization of results to entities that
were not observed. An external true clustering is usually not 
available in reality. Using it for assessment of 
clustering quality where it exists may not help much to 
clarify the characteristics of the clustering methods. Known ``true'' 
classes in datasets where they exist may
deviate systematically from unknown classes of interest in real clustering 
problems. Definitions 
based on probability models suffer from instability. Sometimes 
a researcher may
have a loss function in mind that formalizes the practical problem, but
often this involves an unobservable truth and cannot be directly computed on 
the data alone, in which case it relies on model assumptions and the 
comments in Section \ref{sdpm} apply. 

In any case, 
researchers may have a more complex informal idea of a cluster in mind 
than what can be captured by a formal definition. The definitions of true
clusters should be taken as helpful constructs that support clarification
and transparent comparison of methods, but they should not be taken
as the ultimate clustering truth. Researchers may also complement
formal definitions by less formal
descriptions of more general cluster shapes they are interested in, for 
example ``our method should find elliptical clusters with light tails that
can reasonably be approximated by Gaussian distributions but are separated well
enough that there is a density valley (depth to be defined) between them''.
Methods can then be compared by distributions that fit this description.
Despite all the shortcomings, it would be a strong progress for scientific
communication to accompany the introduction of new clustering
methods regularly with an explicit definition of the clustering problem. 

\section{Implications for cluster analysis research and practice}\label{sprac}
\subsection{Choice of a clustering method in practice}\label{schoice}
If researchers want to find true or real clusters, they have to specify 
what kind of truth they are interested in and what should constitute a ``real'' 
cluster. An appropriate clustering method can be found by connecting the
characteristics of the clustering method to what is desired according to the
researchers' cluster concept. Some methods optimize certain
characteristics directly (such as $k$-means for representing cluster members
by centroids), and in further cases experience
and research suggest typical behavior ($k$-means tends to produce clusters of
roughly equal size and spherical shape, 
whereas methods looking for high-density areas may produce
clusters of very variable size and shape). 
Other characteristics such as stability are
not involved in the definition of most clustering methods, but can be used to
validate clusterings and to compare clusterings from different methods by use
of resampling techniques (\cite{TiWa05}). 
Realist clustering aims can often be related to 
desirable characteristics that can be computed from the data. A more direct 
approach to method choice for realist clustering aims is possible if the 
researchers can specify a probability model and a formal definition of 
truth for the problem under study. Methods with good statistical properties 
for estimating this truth qualify for being chosen, preferably if they can
still do a good job if the model assumptions
are slightly violated. Even realist clustering is a constructive act in the 
sense that
the researchers need to construct their concept of ``real/true'' 
clusters, and in the interest of scientific communication it is desirable
to make this explicit. 

The task of choosing a clustering method is made harder by the fact that in
many applications more than one of the listed characteristics is relevant.
Clusterings may be used for several purposes, and desired characteristics may 
not be well defined, e.g.,  in exploratory data analysis, or in cases 
where the connection between the interpretation of
the clusters and the data is rather loose. 

The specification of a cluster concept that captures a researcher's informal
idea of true clusters is a hard problem, too. Often researchers only
find out that their initial specification was not appropriate if they see
what clustering this yields from their data. I have come across 
such situations often in advisory work. E.g., researchers may 
realize that the used methodology needs to enforce the connection of their 
clustering to an external variable to which their clustering should be 
related, but which they did not specify initially because they believed that 
this
would happen automatically. Or they realize that small clusters are useless for
them only after finding out that
their initially preferred method produces such 
small clusters in their data. 
This illustrates the value of active scientific realism as
complement to constructivism (and the value of cluster validation); 
the researcher's constructs are required, but 
the researchers should be open to change them responding to input from 
the reality outside their control.

\subsection{Comparison of clustering methods}
Although in reality the choice of a clustering method needs to depend on the
context and the clustering aim, research comparing clustering methods 
independently of specific applications is useful because it adds to the 
understanding of the characteristics of the clustering methods. However, as
mentioned in Section \ref{sdefdata} already, in most published 
comparisons of clustering methods 
the authors seem to be far to keen to 
produce simple rankings of methods without providing any insight regarding
what can be learned about the suitability of different methods for different
clustering aims. I have hardly seen any study in which different 
clusterings of the same data or of data from the same probability 
model have been treated as legitimate and were used to tell the 
implicit cluster concepts of different models apart
(\cite{Hennig10merge,ABDBL12,ABDSL13} are examples where this is done). 
Characteristics such as those listed in
Section \ref{sdesir} could be used to evaluate what clustering methods do 
best according to various different characteristics 
datasets without given truth, and they 
could also be used to characterize the true classes in situations where these
classes 
are given, which could help to understand more precisely 
what can be learned from the performance in these cases. Mixture models with
a range of true parameters and component distributions are
occasionally used in comparative studies in a slightly more pluralist way 
with the result that different methods ``win'' different mixtures, although
usually without 
questioning the idea that there is only one true clustering for any 
fixed choice of mixture parameters. Looking at various fixed sets of 
parameters and distributions is more informative for understanding the 
methods in detail than aggregating simulations
with randomly chosen parameters, as some authors seem to prefer, probably 
because this approach can generate a single ranking of methods out of 
many different models.

\subsection{Context-driven vs. data-driven decision making} 
There are a number of other decisions that have to be made when carrying
out a cluster analysis, such as standardization and transformation of 
variables, definition of a dissimilarity measure etc. Similar considerations
as before 
apply regarding the idea that there is a single ``best'' way of
doing this, and their dependence on the context and the clustering aim. 
A number of these decisions is discussed in \cite{HenLi13}.

Here is an exemplary remark regarding variable selection and dimension 
reduction. Many methods are currently advertised for performing this task
automatically. Often they are motivated by their performance in probability
models with a few truly informative and some further homogeneous ``noise''
variables (often
following a Gaussian or uniform distribution). These models capture the idea 
that indeed some variables are relevant for 
clustering and some others are not, abstracted from the meaning of these 
variables. But in real applications, in which the variables have a meaning
that is of substantial importance for the clustering task, choosing different
variables changes the meaning of the resulting clustering. E.g., in a
dataset of students with marks on a number of courses and some standard 
socio-demographic information, one may be interested for different reasons
in clusterings of the marks from science courses, those from 
humanities courses, all courses combined, the socio-demographic information,
or all information combined. It cannot be decided by automatic techniques 
in which of these clusterings the researchers should be interested, and
whether certain variables ``do not cluster'' and whether they then
should not be involved in the computation of the clustering of interest
depends on the context and the clustering aims. 

Regarding the choice of a dissimilarity measure, consider again
the example of data on a central ball and data on a separated ring around it. 
In Section \ref{sext} it was mentioned that 2-means (based on Euclidean data) partitions such a dataset in a way different from ball vs. ring. Assuming that ball vs. ring is the correct partition, one could argue that one should use a different, data driven, dissimilarity (e.g., a path-baed distance) for such data. But if both the Euclidean distance and the use of 2-means have a context-driven justifcation, it is more appropriate to question the intuitive assumption about what the correct partition is. 

\section{Conclusion}

It seems to me that a misguided 
desire for uniqueness and context-independent objectivity
makes many researchers reluctant to
specify desired characteristics and choose a clustering method accordingly,
because they hope that there is a universally optimal method that will just
produce ``natural'' clusters.
Probably for such reasons there is currently only very little
research investigating the characteristics of
methods in terms of the various cluster characteristics that could be of 
interest in different applications of clustering. Also probably 
many researchers are worried about the fact that too strong subjective 
impact could bias analyses and conclusions and could violate the principles of
science because
it will yield results that clearly depend on the observer, see Section 
\ref{sconstr}. 

As pointed out before, there is a tension between the 
scientific goal of general agreement and the acknowledgment of 
individual differences and the unavoidable impact of the individual's point 
of view. Indeed it is important that individual decisions and their 
rationale are made transparent, 
and that they are made in such a way that the ``reality outside our control''
still can deliver its message. E.g.,, variables should be chosen, 
because they are relevant for the 
research question of interest, and not because they produce a specific 
clustering that the researcher wants to promote for some reason. 
There are a number of reasons to make decisions in a data dependent manner,
particularly if the initial analysis of the data reveals that the researchers 
did not properly formalize their aims (see Section \ref{schoice}), in which
case a confirmation on new data (or left out validation data) 
without making data dependent decisions will
normally be required to convince the audience that the results are meaningful.

The philosophical perspective presented here tries to explain how cluster 
analysis can at the same time be strongly dependent on contexts, aims and 
decisions of the researcher, but also scientific, transparent and clear
regarding its underlying concepts and aims, and open to impact 
from Chang's reality outside our control.

I think that the general philosophical considerations apply to much wider areas
of statistics and data analysis; in cluster analysis the plurality of 
definitions, approaches and ideas of truth is particularly striking and 
better visible than elsewhere, but believing in a unique ``natural'' truth
has problematic implications elsewhere as well.

{\bf Acknowledgement:} This work is supported by 
EPSRC grant EP/K033972/1.

\bibliographystyle{chicago}
\bibliography{trueclusters}

\end{document}